\documentclass[twocolumn,showpacs,preprintnumbers,amsmath,amssymb]{revtex4}

\usepackage{graphicx}
\usepackage{dcolumn}
\usepackage{bm}

\def\be{\begin{equation}}
\def\ee{\end{equation}}
\def\bes{\begin{equation*}}
\def\ees{\end{equation*}}
\def\bea{\begin{eqnarray}}
\def\eea{\end{eqnarray}}
\def\beas{\begin{eqnarray*}}
\def\eeas{\end{eqnarray*}}

\begin{document}

\title{Exact Groundstates of Rotating Bose Gases close to a Feshbach Resonance}

\author{N. R. Cooper} 

\affiliation{Cavendish Laboratory, Madingley Road, Cambridge, CB3 0HE,
United Kingdom.}

\date{August 14, 2003}

\begin{abstract}

We study the groundstates of rotating Bose gases when interactions are
affected by a nearby Feshbach resonance.  We show that exact
groundstates at high angular momentum can be found analytically for a
general and realistic model for the resonant interactions.  We
identify parameter regimes where the exact groundstates are exotic
fractional quantum Hall states, the excitations of which obey
non-abelian exchange statistics.

\end{abstract}


\maketitle

Dilute atomic gases are fascinating systems with which to investigate
the unusual properties of Bose-condensed systems when subjected to
rotation. They offer the possibility of entering a regime of high
rotation rate where novel uncondensed groundstates can
appear\cite{wgs}.  At sufficiently high angular momentum, the
groundstate of a system of bosonic atoms interacting by contact
repulsion is the bosonic Laughlin state\cite{wgs,laughlinstates}. In
addition, numerical studies \cite{wg,cwg,RegnaultJ03} have indicated
that as the angular momentum is increased many other
strongly-correlated groundstates appear before the Laughlin state is
reached.

In order to facilitate the experimental observation of the
strongly-correlated groundstates, it has been proposed to use a
Feshbach resonance to increase the energy scale of two-body
interactions\cite{BhongaleMH03}. In addition, the Feshbach
resonance induces an energy dependence to the two-body scattering
amplitude and leads to a non-zero concentration of bound
molecular states of the atoms\cite{TimmermansTCHK99,HollandPW01}. It
is therefore possible that new strongly-correlated groundstates may
form\cite{BhongaleMH03,FischerFR02}.

In this paper we show that {\it exact} groundstates at high angular
momentum can be found for a general and realistic model of atoms
interacting via a Feshbach resonance.  We show that the Laughlin state
of atoms is the exact groundstate of this model over a range of
parameter space. Furthermore, we identify parameter regimes where the
exact groundstates are strongly-correlated atom/molecule mixtures, which are
directly related to the unconventional fractional quantum Hall states
proposed by Moore and Read\cite{MooreR91} and by Read and
Rezayi\cite{ReadR99} (sometimes referred to as the ``Pfaffian'' and
the ``parafermion'' states). These many-body states have very exotic
physical properties, including non-abelian exchange statistics of
their quasiparticle excitations\cite{MooreR91,ReadR99}.  
Exact results are extremely rare in the field of strongly-correlated
many-body systems, and provide a crucial underpinning to
its study.
Our results establish the importance of these exotic
strongly-correlated states for rotating atom gases in a realistic
physical situation.

We consider a system of bosonic atoms and molecules confined to a
harmonic trap with cylindrical symmetry about the $z$-axis.  We denote
the natural frequencies of the trap by $\omega_\parallel$ and
$\omega_\perp$ in the axial and transverse directions.  The
Hamiltonian we consider is the same as that used to explore the
effects of a Feshbach resonance in other
contexts\cite{TimmermansTCHK99,HollandPW01}.
We write the Hamiltonian in the non-rotating frame in the form 
\be
\label{eq:split} 
\hat{H} =  \hat{H}_K + \hat{H}_F + \hat{H}_I+
 (\hbar\omega_\perp+\hbar\omega_\parallel/2)\hat{N}
+  \hbar\omega_\perp \hat{L}
\ee
defining
\begin{eqnarray*}
 \hat{H}_K & \equiv & \int
\hat{a}^\dag_{\bm{r}}\left[ \hat{h}_a - (\hbar\omega_\perp + \hbar\omega_\parallel/2)\right]
\hat{a}_{\bm{r}} \;d^3\bm{r} \\
 & & + \; \hat{m}^\dag_{\bm{r}}\left[ \hat{h}_m - (\hbar\omega_\perp + \hbar\omega_\parallel/2)\right]
\hat{m}_{\bm{r}} \;d^3\bm{r} \\ 
\hat{H}_F
& \equiv & \Delta
\int
\hat{m}^\dag_{\bm{r}}\hat{m}_{\bm{r}} \;d^3\bm{r}  + \frac{U_{aa}}{2} \int
\hat{a}^\dag_{\bm{r}} \hat{a}^\dag_{\bm{r}} \hat{a}_{\bm{r}}\hat{a}_{\bm{r}} \;d^3\bm{r}\\
 & & + \frac{g}{\sqrt{2}} \int \left[
\hat{m}^\dag_{\bm{r}} \hat{a}_{\bm{r}} \hat{a}_{\bm{r}} + \hat{m}_{\bm{r}} \hat{a}^\dag_{\bm{r}}
\hat{a}^\dag_{\bm{r}} \right] \;d^3\bm{r} \\
\hat{H}_I &
\equiv &  U_{am} \!\!\int \!\!\hat{m}^\dag_{\bm{r}}\hat{a}^\dag_{\bm{r}} \hat{a}_{\bm{r}} \hat{m}_{\bm{r}} \;d^3\bm{r} +  \frac{U_{mm}}{2} \!\!\int \!\hat{m}^\dag_{\bm{r}} \hat{m}^\dag_{\bm{r}} \hat{m}_{\bm{r}}
\hat{m}_{\bm{r}} \;d^3\bm{r} 
\end{eqnarray*}
where $\hat{a}^{(\dag)}_{\bm{r}}$ and $\hat{m}^{(\dag)}_{\bm{r}}$ are
the (bosonic) field operators for atoms and molecules.

$\hat{H}_K$ is the kinetic energy (relative to the zero-point energy)
in a frame rotating at $\omega_\perp$ about the $\hat{z}$-axis, with
\begin{eqnarray*}
\hat{h}_a & = & \frac{1}{2M} \left[-i\hbar\bm{\nabla} - \bm{A}_a(\bm{r})\right]^2 
+ \frac{1}{2} M\omega_\parallel^2 z^2 \\
\hat{h}_m & = & \frac{1}{4M} \left[-i\hbar\bm{\nabla} - \bm{A}_m(\bm{r})\right]^2 
+ \frac{1}{2} (2M)\omega_\parallel^2 z^2 
\end{eqnarray*}
where $\bm{A}_a = M \omega_\perp \hat{\bm{z}}\times\bm{r}$ and
$\bm{A}_m = 2M \omega_\perp \hat{\bm{z}}\times\bm{r}$.  The
eigenstates of $\hat{h}_a$ and $\hat{h}_m$ are the products of
symmetric-gauge Landau level states (in the $xy$-coordinates) and
simple harmonic oscillator states (in $z$).  The single-particle
states for atoms and molecules differ only in the natural quantum
lengthscales: in the Landau level states the ``magnetic lengths'' for
the atoms and molecules are $\ell_a = \sqrt{\hbar/(2M\omega_\perp)}$
and $\ell_m=\ell_a/\sqrt{2}$; in the simple harmonic oscillator
states, the ``oscillator lengths'' are $w_a =
\sqrt{\hbar/(M\omega_\parallel)}$ and $w_m=w_a/\sqrt{2}$. (In each
case, the $1/\sqrt{2}$ reduction is due to the molecule/atom mass
ratio of $2$.)

$\hat{H}_F$ describes atom-atom scattering via the Feshbach
resonance\cite{TimmermansTCHK99}: $g$ is the strength of
hybridisation; the detuning $\Delta$ is the energy difference between
the lowest energy state of one molecule and the lowest energy state of
two atoms in the rotating frame\cite{detune}, which can be tuned
continuously with an applied magnetic field.  $\hat{H}_I$ describes
elastic atom-molecule and molecule-molecule scattering.
We shall consider only the case of repulsive bare interactions
$U_{aa},U_{am}, U_{mm} \geq 0$ and positive detuning $\Delta > 0$.
However, note that we make no assumptions concerning the relative sizes of the
interactions and the level spacings of the confinement potential.

The Hamiltonian commutes with both the total number of nuclei (in
atoms and molecules) \bes \hat{N} \equiv \int
\left[\hat{a}^\dag_{\bm{r}} \hat{a}_{\bm{r}} + 2
\hat{m}^\dag_{\bm{r}} \hat{m}_{\bm{r}}\right] d^3\bm{r}
\ees 
and the total angular momentum (in units of $\hbar$)
\bes 
\hat{L} \equiv
\hat{\bm{z}}\cdot\int
\left[\hat{a}^\dag_{\bm{r}}\left(i\bm{\nabla}\times\bm{r}
\right)\hat{a}_{\bm{r}} + \hat{m}^\dag_{\bm{r}}
\left(i\bm{\nabla}\times\bm{r} \right)\hat{m}_{\bm{r}}\right] d^3\bm{r}
\;.
\ees
We shall look for the groundstates of $\hat{H}$ at fixed $N$ and
$L$. The terms in (\ref{eq:split}) proportional to $N$ and $L$ are
then just constant energy offsets, and we need only consider the
action of $\hat{H}_K$, $\hat{H}_F$ and $\hat{H}_I$.

The crucial parameter controlling the nature of the groundstates of a
system of atoms at large angular momentum ($L\gg N\gg 1$) is the
filling fraction\cite{cwg}.  In the present situation of an
atom/molecule mixture, we define the {\it total filling fraction} in
terms of $N$ and $L$ by
$$\nu_T \equiv \frac{N^2}{2L}\;.$$ Denoting the mean 2D density of
nuclei in both atoms and molecules by $n_T = \bar{n}_a + 2\bar{n}_m$,
where $\bar{n}_{a,m}$ are the expectation values of the atomic and
molecular 2D densities, then the total filling fraction may be written
$$\nu_T = n_T(2\pi\ell_a^2) = \bar{\nu}_a + 4\bar{\nu}_m$$
where the individual filling fractions\cite{cwg} for the atoms and
molecules alone are $\bar{\nu}_a \equiv \bar{n}_a (2\pi\ell_a^2)$ and
$\bar{\nu}_m \equiv \bar{n}_m(2\pi\ell_m^2)$.  The factor of $4$
appearing in this formula is a common feature of quantum Hall states
containing bound pairs \cite{Halperin83}.

We now turn to look for the exact groundstates of the Hamiltonian at
fixed total $N$ and $L$.  Substantial progress can be made by noting
that, owing to the contact nature of the atom-atom interaction and the
atom-molecule hybridisation, the atomic field operators in $\hat{H}_F$
can be eliminated in favour of an atom pair operator
$$\hat{P}^{(\dag)}_{\bm{r}} \equiv 
\hat{a}^{(\dag)}_{\bm{r}}\hat{a}^{(\dag)}_{\bm{r}} \;.$$ 
The resulting quadratic form can be diagonalised to give
\begin{eqnarray*}
\hat{H}_F & = & \int \left\{
 \frac{\lambda_-}{{\alpha^2+\beta^2}} \left(\alpha \hat{m}^\dag_{\bm{r}} +
 \beta \hat{P}^\dag_{\bm{r}}\right) \left(\alpha \hat{m}_{\bm{r}} + \beta
 \hat{P}_{\bm{r}}\right)\right. \\ & & \left. +
 \frac{\lambda_+}{{\alpha^2+\beta^2}} \left(\beta \hat{m}^\dag_{\bm{r}} -
 \alpha \hat{P}^\dag_{\bm{r}}\right) \left(\beta \hat{m}_{\bm{r}} - \alpha
 \hat{P}_{\bm{r}}\right)\right\} d^3\bm{r}
\end{eqnarray*}
where
\begin{eqnarray*}
\lambda_\pm & = & \frac{1}{2}\left( \Delta + U_{aa}/2 \pm \sqrt{(\Delta-U_{aa}/2)^2+2g^2}\right) \\
\alpha & = & g/\sqrt{2}\\
\beta & = & \lambda_- -\Delta \;.
\end{eqnarray*}
$\hat{H}_F$ is positive semi-definite provided $\lambda_-\geq 0$,
which is equivalent to the condition
\begin{equation}
\label{eq:inequality}
\frac{g^2}{\Delta} \leq U_{aa} \;.
\end{equation}
Under this condition the reduction of energy of a pair of atoms arising from hybridisation with the molecular level is less than
the energy cost from the repulsive contact interaction to bring the
two atoms together: there remains a net repulsion between pairs of
atoms.

We shall identify exact groundstates of the Hamiltonian, under the
condition (\ref{eq:inequality}).  Our strategy is to construct
many-particle states (for a given $N$ and $L$) which are simultaneous
zero-energy eigenstates of $\hat{H}_K$, $\hat{H}_F$, and
$\hat{H}_I$. Since each of these terms is positive semi-definite, the
resulting states are guaranteed to be groundstates of the total
Hamiltonian (\ref{eq:split}) for given $N$ and $L$.

We start from a {\it purely atomic state} $|\Psi\rangle_a$ (for which
$\hat{m}_{\bm{r}} | \Psi\rangle_a = 0$), and which is  an eigenstate
of $\hat{N}$ and $\hat{L}$. Consider the (unnormalised) state
$|\Psi\rangle$ formed by
\begin{eqnarray}
\label{eq:rotated}
|\Psi\rangle & = & \hat{R} | \Psi\rangle_a\\ \hat{R} & \equiv &
\exp\left(\frac{\alpha}{\beta} \int \hat{P}_{\bm{r}}\hat{m}^\dag_{\bm{r}}d^3\bm{r}
\right) \;.
\end{eqnarray}
Since $\hat{R}$ commutes with both $\hat{N}$ and $\hat{L}$, the number
of nuclei and the total angular momentum of $|\Psi\rangle$ are both
equal to those of the underlying atomic state $|\Psi\rangle_a$.

The operator $\hat{R}$ induces hybridisation of pairs of atoms in
$|\Psi\rangle_a$ into molecules, in such a way that the new state
(\ref{eq:rotated}) is annihilated by the term containing
$\lambda_+$ in $\hat{H}_F$. This follows from noting that $[\hat{m}_{\bm{r}},
\hat{R}] = (\alpha/\beta) \hat{R} \hat{P}_{\bm{r}}$, so
$$\left(\beta \hat{m}_{\bm{r}} - \alpha \hat{P}_{\bm{r}}\right) \hat{R}|\Psi\rangle_a    =  \hat{R} \left(\beta \hat{m}_{\bm{r}} \right) |\Psi\rangle_a    
  = 0$$
since $\hat{m}_{\bm{r}}|\Psi\rangle_a = 0$.
We can describe the residual effects of $\hat{H}_F$ on the states
(\ref{eq:rotated}) by deriving the action of the operator
$\hat{R}^\dag\hat{H}_F\hat{R}$ within the subspace of purely atomic
states. Using $\hat{m}_{\bm{r}}|\Psi\rangle_a =0$, we find
\begin{equation}
\label{eq:rtwo}
\hat{R}^\dag \hat{H}_F \hat{R} = \lambda_-\left(1+\frac{\alpha^2}{\beta^2}\right)\int \hat{a}^\dag_{\bm{r}} \hat{a}^\dag_{\bm{r}} \hat{O}
\hat{a}_{\bm{r}}\hat{a}_{\bm{r}}\;d^3\bm{r} \;.
\end{equation}
where $\hat{O}\equiv \;\;: \exp\left(\frac{\alpha^2}{\beta^2} \int
\hat{P}^\dag_{\bm{r}}\hat{P}_{\bm{r}}d^3\bm{r}\right) \!:$ (the colons
denote normal ordering).  We shall refer to the operator
(\ref{eq:rtwo}) as an ``effective atomic Hamiltonian''. We emphasise
that it is not a unitary transformation of $\hat{H}_F$. However, its
spectrum does allow one to deduce all eigenstates of $\hat{H}_F$ which
are of the form (\ref{eq:rotated}), which is sufficient for our
purposes.  The residual interaction is an effective two-body contact
repulsion between the underlying atoms (with a strength reduced from
$U_{aa}/2$ by the hybridisation).

\begin{figure}
\includegraphics{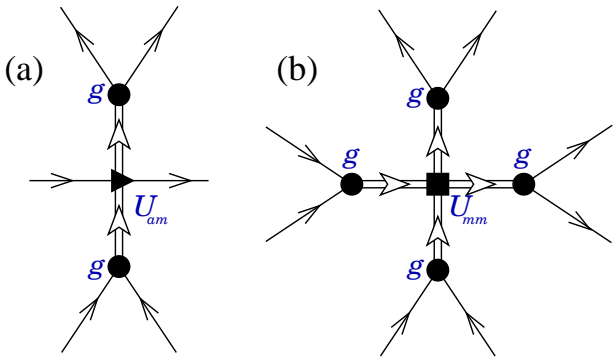}
\caption{\label{fig:feynman} Processes that generate effective (a)
three-body and (b) four-body interactions between the atoms.  A single
(double) line denotes the propagator for an atom (molecule).}
\end{figure}

Similarly, the ``effective atomic Hamiltonian''
from $\hat{H}_I$ is
\begin{eqnarray}
\label{eq:rthree}
 \hat{R}^\dag \hat{H}_I \hat{R}& = &
\left(\frac{\alpha}{\beta}\right)^2U_{am} \int
\left[\hat{a}^\dag_{\bm{r}}\right]^3  \hat{O}\left[\hat{a}_{\bm{r}}\right]^3\;
d^3\bm{r}\\ & + &
\label{eq:rfour}
 \left(\frac{\alpha}{\beta}\right)^4\frac{U_{mm}}{2}
\int \left[\hat{a}^\dag_{\bm{r}}\right]^4 \hat{O}
\left[\hat{a}_{\bm{r}}\right]^4\; d^3\bm{r} \;.
\end{eqnarray}
The atom-molecule and molecule-molecule interactions generate
effective three- and four-body contact interactions for the underlying
atoms. These arise from the processes indicated schematically in
Fig.~\ref{fig:feynman}.

We now consider the action of the kinetic energy operator $\hat{H}_K$
on the states (\ref{eq:rotated}).  It is helpful to expand the field
operators in the eigenstates of $\hat{h}_{a,m}$: we label these states
by the quantum numbers $n,m$ and $q$, where $n=0,1,2,\ldots$ and
$m=-n,-n+1,\ldots$ are the Landau level index and the angular momentum
quantum number (describing motion in the $xy$-plane), and $q=0,1,2,\ldots$
is the oscillator index of the axial motion.  We then define
$\hat{a}^\dag_{nmq} \equiv \int d^3\bm{r} \; \hat{a}^\dag_{\bm{r}}
\langle \bm{r} | n,m,q\rangle_a$, and $\hat{m}^\dag_{nmq} \equiv \int
d^3\bm{r} \; \hat{m}^\dag_{\bm{r}} \langle \bm{r} | n,m,q\rangle_m$.
The states $|n,m,q\rangle_m$ and the states $|n,m,q\rangle_a$ are each
assumed to provide a complete basis in which arbitrary wavefunctions
can be expanded (over all lengthscales for which the contact
interaction Hamiltonian is assumed valid).  We may then write \be
\label{eq:equiv}
\int \hat{P}_{\bm{r}}m^\dag_{\bm{r}}  \; d^3\bm{r}   = \sum_{n,m,q} \hat{P}_{nmq}m^\dag_{nmq}
\end{equation}
where
\begin{widetext}
\be
\label{eq:pdag} \hat{P}^\dag_{nmq} \equiv  \sum_{n_1,m_1,q_1}\sum_{n_2,m_2,q_2} \left\{\hat{a}^\dag_{n_1 m_1 q_1}  \hat{a}^\dag_{n_2 m_2 q_2} 
 \int  { }_a\langle  n_1,m_1,q_1|\bm{r}\rangle   { }_a\langle  n_2,m_2,q_2|\bm{r}\rangle\langle \bm{r} | n,m,q\rangle_m \;d^3\bm{r}\right\}
\ee
\end{widetext}
We now insist that {\it all atoms in the state $|\Psi\rangle_a$
are in the lowest Landau level and lowest axial oscillator state}:
that is, in states with $n=q=0$.  In this case, the ``effective atomic
Hamiltonian'' $\hat{R}^\dag \hat{H}_K \hat{R}$ vanishes identically.
This surprising result follows from the fact that the $n=q=0$
wavefunctions for a molecule, $\langle \bm{r}|0,m,0\rangle_m$, which
are proportional to
$$(x+iy)^m e^{-(x^2+y^2)/4\ell_m^2}e^{-z^2/2w_m^2}$$ provide a
complete set of states for  the product of any two atomic wavefunctions
with $n=q=0$,  $\langle \bm{r}| 0,m_1,0\rangle_a \langle
\bm{r}| 0,m_2,0\rangle_a$, which is proportional to
$$(x+iy)^{m_1+m_2} e^{-2(x^2+y^2)/4\ell_a^2}e^{-2z^2/2w_a^2}\;.$$ The
completeness relation holds since $m$ can span all possible values of
$m_1+m_2$ and since the arguments in the exponentials are identical
due to $\ell_m = \ell_a/\sqrt{2}$ and $w_m =
w_a/\sqrt{2}$. Consequently, from (\ref{eq:equiv},\ref{eq:pdag}), the
action of $\int \hat{P}_{\bm{r}}\hat{m}^\dag_{\bm{r}} d^3\bm{r}$ on
any state in which all atoms have $n=q=0$ is equivalent to the action
of $\sum_{m} \hat{P}_{0m0}\hat{m}^\dag_{0m0}$. In the new state
(\ref{eq:rotated}) molecules are only in the lowest Landau level and
lowest axial oscillator state ($n=q=0$): the state (\ref{eq:rotated})
is therefore annihilated by $\hat{H}_K$. We emphasise that choosing to
construct $|\Psi\rangle_a$ from states in which atoms are only in
states with $n=q=0$ does {\it not} impose further restrictions on the
parameters in the model (\ref{eq:split}); we make this choice because
the exact groundstates are of this form in the situations we study.

We are now in a position to deduce exact groundstates of the
Hamiltonian (\ref{eq:split}) at particular values of $N$ and $L$.
States of the form (\ref{eq:rotated}) based on an underlying atomic
state $|\Psi\rangle_a$ in which all atoms have $n=q=0$ are zero-energy
eigenstates of $\hat{H}_K$.  To identify zero-energy eigenstates of
$\hat{H}_F$ and $\hat{H}_I$, we further require that the underlying
atomic wavefunction $|\Psi\rangle_a$ be annihilated by the effective
two-, three-, and four-body contact interactions
(\ref{eq:rtwo},\ref{eq:rthree},\ref{eq:rfour}).

Zero-energy eigenstates of the two-body contact interaction (and
therefore also of the three- and four-body interactions), can be
constructed in the lowest Landau level for any $L\geq N(N-1)$ by
choosing $|\Psi\rangle_a$ to be the bosonic Laughlin state (at
$L=N(N-1)$), or the Laughlin state with additional quasiholes (at
$L>N(N-1)$).  The Laughlin ket state of $N$ atoms, $|\Psi_L\rangle_a$,
may be constructed from its first quantised wavefunction by \bea
\label{eq:laughlin}
|\Psi_L\rangle_a & = & \prod_{i=1}^N \int\!\!  d^3\bm{r}_i
\Psi_L(\{\bm{r}_i\}) \prod_{j=1}^N \hat{a}^\dag_{\bm{r}_j} |0\rangle\\
\nonumber
\Psi_L(\{\bm{r}_i\}) & = & \prod_{i<j=1}^N (\eta_i-\eta_j)^2
e^{-\sum_i \left[\frac{|\eta_i|^2}{4\ell_a^2} +
\frac{z_i^2}{2w_a^2}\right]} \eea where $\eta_i \equiv (x_i+iy_i)$
and $|0\rangle$ is the particle vacuum.  (We omit normalisation
factors.)

For $L=N(N-1)$ the Laughlin state is the only state in the lowest
Landau level that is a zero-energy eigenstate of the two-body contact
interaction.  The state $|\Psi_L\rangle = \hat{R} |\Psi_L\rangle_a$ is
therefore the non-degenerate groundstate of the model Hamiltonian
(\ref{eq:split}) at $L=N(N-1)$, {\it i.e.} at total filling fraction
$\nu_T = 1/2$.  Note, however, that since $|\Psi_L\rangle_a$ has no
amplitude for two atoms to coincide, consequently $\hat{P}_r
|\Psi_L\rangle_a=0$ and $|\Psi_L\rangle =\hat{R} |\Psi_L\rangle_a =
|\Psi_L\rangle_a$. {\it The exact groundstate at $\nu_T=1/2$ is the
bosonic Laughlin state for atoms (\ref{eq:laughlin}).}  We emphasise
that this result is valid for arbitrarily strong atom-atom,
atom-molecule and molecule-molecule interactions.  The only
restrictions we impose are that $g^2/\Delta < U_{aa}$, and $\Delta,
U_{aa}, U_{am}, U_{mm}\geq 0$.  The pairing correlations between atoms
suggested by Bhongale {\it et al.}\cite{BhongaleMH03} do not appear in
the groundstate under these conditions.

It is particularly interesting to consider tuning the parameters (for
example by varying the detuning $\Delta$) such that $g^2/\Delta =
U_{aa}$. Then $\lambda_-=0$ and the effective two-body repulsion
(\ref{eq:rtwo}) vanishes: the effective atomic Hamiltonian consists
only of the repulsive three- and four-body contact interactions
(\ref{eq:rthree},\ref{eq:rfour}).

Exact zero-energy eigenstates of these interactions can be constructed
in the lowest Landau level for $L \geq N(N/2-1)$, from the Moore-Read
(``Pfaffian'') state\cite{MooreR91}.  The Moore-Read state is the
unique zero-energy eigenstate of a three-body contact interaction at
$L=N(N/2-1)$\cite{GreiterWW91}. Hence, we establish that
$|\Psi_{MR}\rangle = \hat{R} |\Psi_{MR}\rangle_a$ is the
non-degenerate groundstate of the Hamiltonian (\ref{eq:split}) at
$L=N(N/2-1)$, which corresponds to $\nu_T = 1$\cite{kets}.

If $U_{am}/U_{mm}\ll g^2/(\Delta^2 w_a \ell_a^2)$, then the
effective three-body contact interaction (\ref{eq:rthree}) is
negligible compared to the effective four-body interaction
(\ref{eq:rfour}) when $\nu_T\sim O(1)$. It is then of
interest to set $U_{am}=0$ and find exact groundstates of the
four-body contact interaction alone (\ref{eq:rfour}).  This can be
achieved in the lowest Landau level for $L\geq N(N/3-1)$ by use of the
$k=3$ Read-Rezayi (``parafermion'') wavefunction\cite{ReadR99}.  From
the uniqueness of this state at $L=N(N/3-1)$\cite{ReadR99}, we
establish that $|\Psi_{k=3}\rangle = \hat{R} |\Psi_{k=3}\rangle_a$ is
the non-degenerate groundstate of (\ref{eq:split}) at $L=N(N/3-1)$,
corresponding to $\nu_T=3/2$.

Note that the exact groundstates which are formed from an underlying
Moore-Read state (at $\nu_T=1$) or $k=3$ Read-Rezayi state (at
$\nu_T=3/2$) are {\it strongly-correlated atom/molecule mixtures}
(since $\hat{P}_r |\Psi_{MR}\rangle_a\neq 0$ and $\hat{P}_r
|\Psi_{k=3}\rangle_a\neq 0$).
These atom/molecule mixtures will
exhibit all the exotic properties that have been predicted for the
Moore-Read and Read-Rezayi states of atoms alone, including
non-abelian statistics for their quasiparticle
excitations\cite{MooreR91,ReadR99}.  Although these states provide
exact descriptions of the groundstates only at $g^2/\Delta = U_{aa}$
(and $U_{am}=0$ for the Read-Rezayi state), since these states have
excitation gaps they should continue to describe the groundstates at
$\nu_T=1$ and $\nu_T=3/2$ over a range of parameter space. Indeed,
these states have been found to provide accurate descriptions of the
groundstates of a purely atomic system ($\Delta\to\infty$) at
$\nu_T=1$\cite{wg,cwg,RegnaultJ03} and $\nu_T=3/2$\cite{cwg}; it is
therefore possible that they provide accurate descriptions of the
groundstates at these filling fractions over the range $g^2/\Delta =
U_{aa}$ to $g^2/\Delta = 0$.

I am grateful to N. Read, S. G. Bhongale and J. N. Milstein for
helpful correspondence.  I acknowledge support from EPSRC grant
no. GR/R99027/01.


\begin{thebibliography}{10}
\bibitem{kets}{The Moore-Read state and $k=3$ Read-Rezayi state for
atoms, $|\Psi _{MR}\rangle_a$ and $|\Psi _{k=3}\rangle_a$, can be
constructed  in the same way as (\ref{eq:laughlin}) from the first quantised wavefunctions given
in\cite{MooreR91,ReadR99}.}
\end{thebibliography}

\end{document}